\documentclass[apj]{emulateapj}

\def\lea{\mathrel{<\kern-1.0em\lower0.9ex\hbox{$\sim$}}}
\def\gea{\mathrel{>\kern-1.0em\lower0.9ex\hbox{$\sim$}}}

\slugcomment{}

\shorttitle{EzGal}
\shortauthors{Mancone, Gonzalez}

\begin{document}


\title{EzGal: A Flexible Interface for Stellar Population Synthesis Models}

\author{Conor L. Mancone and Anthony H. Gonzalez}
\affil{Department of Astronomy, University of Florida, Gainesville, FL 32611}
\email{cmancone@astro.ufl.edu, anthony@astro.ufl.edu}

\begin{abstract}
We present {\itshape EzGal}, a flexible python program designed to easily generate observable parameters (magnitudes, colors, mass-to-light ratios) for arbitrary input stellar population synthesis (SPS) models.  As has been demonstrated by various authors, for many applications the choice of input SPS models can be a significant source of systematic uncertainty.  A key strength of {\itshape EzGal} is that it enables simple, direct comparison of different models sets so that the uncertainty introduced by choice of model set can be quantified.  Its ability to work with new models will allow {\itshape EzGal} to remain useful as SPS modeling evolves to keep up with the latest research (such as varying IMFS; \citealt{cappellari12}).  {\itshape EzGal} is also capable of generating composite stellar population models (CSPs) for arbitrary input star formation histories and reddening laws, and can be used to interpolate between metallicities for a given model set. To facilitate use, we have created a web interface to run EzGal and quickly generate magnitude and mass-to-light ratio predictions for a variety of star formation histories and model sets.  We make many commonly used SPS models available from the web interface including the canonical \citet{BC03} models, an updated version of these models, the Maraston models, the BaSTI models, and finally the FSPS models.  We use {\itshape EzGal} to compare magnitude predictions for the model sets as a function of wavelength, age, metallicity, and star formation history.  From this comparison we quickly recover the well-known result that the models agree best in the optical for old, solar metallicity models, with differences at the $\sim0.1$ magnitude level. Similarly, the most problematic regime for SPS modeling is for young ages ($\la 2$ Gyrs) and long wavelengths ($\lambda \gtrsim 7500$\AA) where thermally pulsating AGB stars are important and scatter between models can vary from 0.3 mags (Sloan i') to 0.7 mags (K$s$).  We find that these differences are not caused by one discrepant model set and should therefore be interpreted as general uncertainties in SPS modeling.  Finally we connect our results to a more physically motivated example by generating CSPs with a star formation history matching the global star formation history of the universe.  We demonstrate that the wavelength and age dependence of SPS model uncertainty translates into a redshift dependent model uncertainty, highlighting the importance of a quantitative understanding of model differences when comparing observations to models as a function of redshift.
\end{abstract}


\defcitealias{CB07}{CB07}
\defcitealias{BC03}{BC03}
\defcitealias{M05}{M05}
\defcitealias{BaSTI}{BaSTI}
\defcitealias{C09}{C09}

\section{Introduction\label{sec:intro}}

\begin{figure*}
{\large \begin{equation}M_{AB}\left[z,t(z,z_f)\right] = -2.5\log\left[ \frac{ \int_{-\infty}^\infty \nu^{-1} (1+z)F_\nu[\nu(1+z),t(z,z_f)] R(\nu) d\nu } {\int_{-\infty}^\infty \nu^{-1} R(\nu) d\nu} \right] - 48.60\label{eq:mags}\end{equation}}
\end{figure*}

Stellar population synthesis (SPS) modeling provides a valuable tool for studying the evolution of a stellar population as a function of time.  For this reason there have been multiple efforts to develop software for modeling the evolution of the spectral energy distribution (SED) of a stellar population.  Because there are a number of unknowns in SPS modeling, such as details of post-main sequence stellar evolution and the form of the initial mass function, models from different groups yield a range of results due to different input ingredients in their models.  A detailed discussion of these uncertainties and their impact can be found in \citet{C09} and \citet{conroy10}.  The net result is that the choice of model set is itself a source of uncertainty when using SPS models.

The use of SPS models is a central ingredient for a wide range of active research programs, as is evident even from a simple literature search.  SPS models are commonly used to perform SED fitting and estimate a diverse set of properties for stellar populations including ages, redshifts, $k$-corrections, and masses (see for example \citealt{blanton07,taylor11,ma12,foto12}).  They are used to fit isochrones to color magnitude diagrams and measure ages and metallicities of resolved stellar populations, to measure the strength of spectral features in observed galaxies, to predict the evolution of a stellar population as a function of age, and to predict observables from simulations (for example \citealt{jonsson06,ata,mancone10,kriek11}).  Because of the utility and ubiquity of SPS models, it is important to have simplifying methods for comparing the models to observations as well as to each other.

Of the many SPS model sets the most commonly used is that of Bruzual \& Chalot (\citetalias{BC03}; 2003) which we use as a reference for comparisons because of its wide use.  Another commonly used model set is that of Maraston (\citetalias{M05}; 2005) which includes a detailed treatment of thermally pulsating AGB (TP-AGB) stars, which can dominate the infrared light of a young stellar population.  An updated treatment of the TP-AGB phase is also incorporated into the latest version of the \citetalias{BC03} models \citep[commonly referred to as CB07]{CB07}.  More recent models include the work of Percival et al. (\citetalias{BaSTI}; 2009) which include not only a range of metallicities but also $\alpha-$enhanced models.  The FSPS models (\citetalias{C09}) from \citet{C09} and \citet{conroy10} are unique in their ability to treat the most important SPS inputs (such as IMF or various uncertain phases of stellar evolution) as free parameters allowing the uncertainties introduced by various SPS inputs to be quantitatively measured.

All of these models predict the evolution of the SED of a stellar population as a function of age, given a star formation history, initial mass function (IMF), and metallicity.  However the easiest to measure observables are not the SED or age, but rather the magnitude and redshift.  Therefore all of these model sets are most useful when they can be easily translated into predictions of magnitude evolution as a function of redshift.  This transformation involves assuming a formation redshift (the redshift at which star formation starts), calculating a cosmology-dependent luminosity distance, and projecting the SEDs through filter response curves to calculate magnitudes, $e$-corrections, and $k$-corrections.  The $e$-corrections specify the amount of observed magnitude evolution that is due to the aging of a stellar population, while the $k$-corrections specify the amount of evolution due to observing a different part of the SED at different redshifts.  Together the $e$-corrections, $k$-corrections, and distance moduli specify the magnitude evolution of a stellar population as a function of redshift.

While these steps are straight-forward, in the past there has not been a simple and consistent tool to do this for all model sets.  \citetalias{BC03} and \citetalias{C09} both come with code for calculating magnitude evolution as a function of redshift and both come with a number of commonly used filter response curves for user convenience.  In contrast, \citetalias{BaSTI} and \citetalias{M05} calculate and distribute the absolute magnitude evolution of the stellar populations for a fixed set of filters.

This lack of directly comparable outputs between different model sets is the reason why we have developed {\itshape EzGal}, a python program that calculates magnitude evolution as a function of redshift from models of the evolution of an SED as a function of age.  {\itshape EzGal} comes with a number of the most commonly used filter response curves, and more can be easily added by the user.  It includes the latest Vega spectrum from STScI\footnote{http://www.stsci.edu/hst/observatory/cdbs/calspec.html} so that magnitudes can be calculated on both the Vega and AB systems.  By using the stellar mass information that comes with all of these model sets {\itshape EzGal} can also calculate mass-to-light ratios in any filter.  This requires calculating the absolute magnitude of the Sun in any filter and so the latest solar spectrum from STScI\footnotemark[2] is also included with {\itshape EzGal}.  {\itshape EzGal} can interpolate between models, which is useful for generating models with the same metallicity from different model sets.  It can also generate CSPs with arbitrary input star formation histories or dust reddening laws.  Finally, {\itshape EzGal} can read in SEDs in ASCII format or in the binary ised format that the \citetalias{BC03} and \citetalias{CB07} models are distributed in.  In principle this allows it to work with any model, enabling easy comparison with any new codes in the future. {\itshape EzGal} is designed to be an easy-to-use tool for predicting observables from SPS models and greatly simplifying the task of comparing different SPS model sets.

This paper explains how {\itshape EzGal} works and gives a detailed comparison between commonly used model sets.  Section \ref{sec:procedure} describes details of how {\itshape EzGal} works and discusses calculating magnitudes (Section \ref{sec:mags}), generating composite stellar populations (Section \ref{sec:csp}), and calculating masses and mass-to-light ratios (Section \ref{sec:mls}).  In Section \ref{sec:compare} we present a detailed comparison between the model sets.  Section \ref{sec:web} lists EzGal resources currently available from the internet such as the web interface.  Our conclusions are found in Section \ref{sec:conclusions}.

\section{Program Procedure}\label{sec:procedure}

\subsection{Calculating Magnitudes}\label{sec:mags}

{\itshape EzGal} calculates apparent magnitudes, absolute magnitudes, $e$-corrections, and $k$-corrections from the model sets as a function of redshift.  Conceptually, these quantities are all easy to calculate and are derived from the rest-frame and observed-frame absolute magnitudes as a function of age and redshift.  {\itshape EzGal} uses Equation \ref{eq:mags} to calculate observed-frame absolute magnitudes as a function of redshift ($z$) and formation redshift ($z_f$).  This equation calculates the absolute AB magnitude as a function of redshift and age, $M_{AB}\left[z,t(z,z_f)\right]$, for an SPS model by projecting the redshifted SED, $F_\nu[\nu(1+z),t(z,z_f)]$, at the given age, $t(z,z_f)$ (with the age determined by redshift and formation redshift), through the filter response curve, $R(\nu)$, and comparing this to the flux of a zero mag AB source.  For the purposes of this equation the SED should have units of ergs$^{-1}$Hz$^{-1}$cm$^{-2}$ and should be the observed flux for a galaxy at a distance of 10 pc.  The age of the galaxy, $t(z,z_f)$, is given by $t(z,z_f) = T_U(z) - T_U(z_f)$ where $T_U(z)$ is the age of the universe as a function of redshift given the cosmology.  By default {\itshape EzGal} assumes a WMAP 7 cosmology \citep[$\Omega_m=0.272, \Omega_\Lambda=0.728, h=0.704$]{komatsu11}, although any cosmology can be used.  To calculate the rest-frame absolute magnitude, {\itshape EzGal} calculates $M_{AB}[ 0, t(z,z_f) ]$.

{\itshape EzGal} also calculates a number of filter properties using standard STScI definitions, including mean wavelength, pivot wavelength, average wavelength, effective dimensionless gaussian width, effective width, equivalent width, and rectangular width.\footnote{http://www.stsci.edu/hst/wfpc2/documents/handbook/cycle17/ch6\_exposuretime2.html\#480221}$^,$\footnote{http://www.stsci.edu/hst/wfc3/documents/handbooks/currentIHB/c06\_uvis06.html\#57}  The conversion from AB to Vega magnitudes is calculated for each filter by using the included Vega spectrum to calculate the AB magnitude of Vega in the filter.  The Vega spectrum is described in \citet{bohlin04} and comes from IUE spectrophotometry from 0.12 - 0.17$\mu$m, HST STIS spectroscopy from 0.17 - 1.01$\mu$m, and a Kurucz model atmosphere at longer wavelengths.  Finally, the absolute magnitude of the Sun is also calculated by projecting the solar spectrum through the filter response curve in the same way as everything else.  The solar spectrum used by EzGal is an observed spectrum of the Sun from 0.12 - 2.5\micron ~\citep{sun} which we have extended using a Kurucz model atmosphere at longer wavelengths.  Specifically, we take a model atmosphere with solar metallicity, $T_{eff}$ = 5777K, and $log_g$ = 4.44, normalize it to match the observed solar spectrum from 1.5 - 2.5 \micron, and then use it where the observed spectrum ends.

\subsection{Calculating Composite Stellar Populations}\label{sec:csp}

{\itshape EzGal} generates composite stellar population (CSP) models from simple stellar population (SSP) models in the standard way.  Conceptually, the SED of a CSP at some age is given by a weighted average of SSPs as a function of age, where the weight for a given SSP is equal to the relative strength of star formation (compared to the total amount of star formation) for the CSP at that time.  The effect of dust can also be included if desired.  Mathematically, {\itshape EzGal} uses Equation \ref{eq:csp} to calculate the evolution of the SED of a CSP as a function of time.

\begin{equation}F(\lambda,t) = \frac{ \int_{0}^t\Psi(t-t')F_{SSP}(\lambda,t')\Gamma(\lambda,t')dt' }{ \int_0^{T_U}\Psi(t')dt' }\label{eq:csp}\end{equation}

In this equation $F(\lambda,t)$ is the flux of the CSP as a function of wavelength and time, $\Psi(t)$ is the star formation rate as a function of time, $F_{SSP}(\lambda,t')$ is the flux of the SSP as a function of wavelength and time, $\Gamma(\lambda,t')$ is the impact of dust as a function of wavelength and time, and $T_U$ is the age of the universe at z=0.  The factor of $\int_0^{T_U}\Psi(t')dt'$ normalizes the CSP such that one solar mass of stars is generated over the entire star forming epoch.  EzGal can work with arbitrary star formation histories and dust laws.  A typical dust law is a \citet{charlot00} dust law with $\Gamma(\lambda) = e^{-\tau(t)(\lambda/5500\mathrm{\AA})^{-0.7}}$ where $\tau(t) = 1.0$ for $t \leq 10^7$yr and $\tau(t) = 0.5$ for $t > 10^7$yr.  Equation \ref{eq:csp} represents the same general methodology used by \citetalias{BC03} and \citetalias{C09} to generate CSPs.  \citetalias{M05} uses a different normalization and instead divides by $\int_0^t\Psi(t')dt'$ so that the CSPs have one solar mass of stars at all ages.  \citetalias{BaSTI} does not provide any CSPs with their models.

In practice {\itshape EzGal} uses Simpson's rule to numerically evaluate the top integral in Equation \ref{eq:csp}.  When performing numeric integration it is often necessary to sub-sample the age grid of the SSPs to properly sample any sharp features in the star formation history or in the evolution of the SEDs.  In order to minimize execution time and still ensure high fidelity in the numeric integration, {\itshape EzGal} uses an iterative algorithm to decide how finely to sub-sample the age grid.  {\itshape EzGal} performs the integral in Equation \ref{eq:csp} at wavelengths of 3000, 8000, and 12000\AA ~with increasingly finer levels of age sub-sampling until the difference between two subsequent integrals drops below some tuneable threshold (in magnitudes).

\begin{deluxetable*}{cccccccc}
\tablecaption{Model Set Properties\label{tbl:models}}
\tablewidth{0pt}
\tablehead{
  \colhead{Name} & \colhead{\# Ages} & \colhead{Metallicity ($Z/Z_\odot$)} & \colhead{$\alpha$-enhanced} & \colhead{\# Metallicities} & \colhead{Salpeter} & \colhead{Chabrier} & \colhead{Kroupa}
}
\startdata
\citetalias{BC03}  & 221 & 0.005 - 2.5 & No  & 6  & Yes & Yes & No  \\
\citetalias{M05}   & 68  &  0.05 - 3.5 & No  & 5  & Yes & No  & Yes \\
\citetalias{CB07}  & 221 & 0.005 - 2.5 & No  & 6  & Yes & Yes & No  \\
\citetalias{BaSTI} & 56  & 0.005 - 2   & Yes & 10 & No  & No  & Yes \\
\citetalias{C09}   & 189 &  0.01 - 1.5 & No  & 22 & Yes & Yes & Yes \\
\enddata
\end{deluxetable*}

\begin{deluxetable*}{ccccc}
\tablecaption{Filter Data\label{tbl:filters}\tablenotemark{a}}
\tablewidth{0pt}
\tablehead{
  \colhead{Name} & \colhead{Pivot Wavelength (\AA)} & \colhead{Rectangular Width (\AA)} & \colhead{M$_\odot$ (AB)} & \colhead{Vega - AB}
}
\startdata
      GALEX FUV            &  1536 &   246 & 17.20 & -2.093 \\
      GALEX NUV            &  2300 &   730 & 10.04 & -1.659 \\
       Sloan u'            &  3556 &   558 &  6.37 & -0.916 \\
  ACS WFC F435W            &  4318 &   845 &  5.37 &  0.102 \\
     WFC3 F438W            &  4325 &   616 &  5.34 &  0.152 \\
       Sloan g'            &  4702 &  1158 &  5.12 &  0.100 \\
  ACS WFC F475W            &  4746 &  1359 &  5.10 &  0.096 \\
     WFC3 F475W            &  4773 &  1343 &  5.08 &  0.096 \\
     WFC3 F555W            &  5308 &  1563 &  4.86 &  0.023 \\
  ACS WFC F555W            &  5360 &  1124 &  4.84 &  0.005 \\
     WFC3 F606W            &  5887 &  2183 &  4.73 & -0.085 \\
  ACS WFC F606W            &  5921 &  1992 &  4.72 & -0.088 \\
       Sloan r'            &  6175 &  1111 &  4.64 & -0.144 \\
     WFC3 F625W            &  6241 &  1461 &  4.64 & -0.150 \\
  ACS WFC F625W            &  6311 &  1308 &  4.63 & -0.165 \\
       Sloan i'            &  7489 &  1045 &  4.53 & -0.357 \\
     WFC3 F775W            &  7647 &  1170 &  4.53 & -0.382 \\
  ACS WFC F775W            &  7691 &  1320 &  4.53 & -0.389 \\
     WFC3 F814W            &  8026 &  1538 &  4.52 & -0.419 \\
  ACS WFC F814W            &  8055 &  1733 &  4.52 & -0.425 \\
       Sloan z'            &  8946 &  1125 &  4.51 & -0.518 \\
 ACS WFC F850lp            &  9013 &  1239 &  4.51 & -0.521 \\
    WFC3 F850lp            &  9167 &  1181 &  4.52 & -0.522 \\
     WFC3 F105W            & 10550 &  2649 &  4.53 & -0.647 \\
     WFC3 F110W            & 11534 &  4430 &  4.54 & -0.761 \\
              J            & 12469 &  2088 &  4.56 & -0.901 \\
     WFC3 F125W            & 12486 &  2845 &  4.56 & -0.903 \\
     WFC3 F140W            & 13922 &  3840 &  4.60 & -1.078 \\
     WFC3 F160W            & 15370 &  2683 &  4.65 & -1.254 \\
              H            & 16448 &  2538 &  4.70 & -1.365 \\
             Ks            & 21623 &  2642 &  5.13 & -1.838 \\
 {\itshape WISE} 3.4$\mu$m & 33682 &  6824 &  5.95 & -2.668 \\
 IRAC 3.6$\mu$m            & 35569 &  6844 &  6.07 & -2.787 \\
 IRAC 4.5$\mu$m            & 45020 &  8707 &  6.57 & -3.260 \\
 {\itshape WISE} 3.6$\mu$m & 46179 & 10508 &  6.62 & -3.307 \\
 IRAC 5.8$\mu$m            & 57450 & 12441 &  7.05 & -3.753 \\
 IRAC 8$\mu$m              & 79156 & 25592 &  7.67 & -4.394 \\
\enddata
\tablenotetext{a}{An up-to-date copy of this table is reproduced at http://www.baryons.org/ezgal/filters.php}
\end{deluxetable*}

We verify our procedure for generating CSPs by comparing magnitude predictions for CSPs generated with {\itshape EzGal} from \citetalias{BC03} and \citetalias{C09} models to magnitude predictions for CSPs generated by the code distributed with \citetalias{BC03} and \citetalias{C09}.  We find differences that are small and negligible: for \citetalias{BC03} the differences in magnitude are $<0.005$ mags for short ($\tau=0.1$Gyr) and long ($\tau=1.0$Gyr) dust-free exponentially-decaying bursts, and for \citetalias{C09} the differences are $<0.01$ mags for short bursts and $<0.005$ mags for long bursts.  These differences are larger than the maximum error set in our numerical integration (0.001 mags), but errors at these levels can easily be accounted for by small differences in the procedures used by different groups.

\subsection{Calculating Mass-to-Light Ratios and Masses}\label{sec:mls}

To calculate rest-frame mass-to-light ratios in any filter, $F$, given redshift and formation redshift, four pieces of information are required: the age as a function of redshift and formation redshift, $t(z,z_f)$, the stellar mass as a function of age, $M_*[t(z,z_f)]$, the rest-frame absolute magnitude of the stellar population as a function of age, $M_F[t(z,z_f)]$, and the absolute magnitude of the Sun in the filter, $M_{\odot,F}$.  Again, the conversion from redshift and formation redshift to age requires assuming a cosmology, for which EzGal defaults to a WMAP 7 cosmology \citep[$\Omega_m=0.272, \Omega_\Lambda=0.728, h=0.704$]{komatsu11}.  The rest-frame mass-to-light ratio in a given filter as a function of redshift and fromation redshift is then given by:

{\large \begin{equation}\frac{M_*}{L_F}(z,z_f) = \frac{M_*[t(z,z_f)]}{ 10^{ -0.4*\{M_F[t(z,z_f)]-M_{\odot,F}\} } }\label{eq:ml}\end{equation}}

{\itshape EzGal} uses Equation \ref{eq:ml} to calculate rest-frame mass-to-light ratios.  It uses its own calculation of the absolute magnitude evolution of a stellar population as a function of age, calculates the absolute magnitude of the Sun using the solar spectrum from STScI, and gets stellar masses directly from the model sets (which typically distribute stellar mass as a function of age along with the SED).  The resulting mass-to-light ratios depend on the chosen model set, star formation history, and initial mass function.  {\itshape EzGal} also calculates an observed-frame mass-to-light ratio as a function of redshift using the observed-frame absolute magnitude of the model and the observed-frame absolute magnitude of the Sun.  The latter is calculated by redshifting the solar spectrum to the given redshift and projecting it through the bandpass normally.

For the purposes of estimating the mass of an observed galaxy only two pieces of information are required from the models: the stellar mass as a function of redshift and the apparent magnitude of the model as a function of redshift.  With these values in hand the mass of an observed galaxy with an assumed redshift ($z$) and formation redshift ($z_f$) can be calculated as:

{\large \begin{equation}M_{*,g}(z,z_f) = M_*(z,z_f)*10^{ -0.4*(m_{g,F} - m_F(z,zf)) }\label{eq:mass},\end{equation}}

\noindent where $M_*(z,z_f)$ is the stellar mass of the model as a function of $z$ and $z_f$, $m_{g,F}$ is the apparent magnitude of the galaxy in a given passband, and $m_F(z,zf)$ is the apparent magnitude of the model in the same passband as a function of $z$ and $z_f$.

\section{Model Comparison}\label{sec:compare}

\subsection{Model Set Overview}

In this paper we compare results from five different SPS model sets: \citetalias{BC03}, \citetalias{M05}, \citetalias{CB07}, \citetalias{BaSTI}, and \citetalias{C09}.  These model sets include a varying range of metallicities and IMFs, and have different spectral resolutions and age grids.  \citetalias{M05} has the highest metallicity model ($Z = 3.5Z_\odot$) while \citetalias{BC03}, \citetalias{CB07}, and \citetalias{BaSTI} have the lowest ($Z = Z_\odot$/200).  \citetalias{C09} has the finest grid in metallicity space with 22 metallicities from $Z = 0.01Z_\odot - 1.5Z_\odot$ and is the only model set that distributes models with all three common IMFs: Salpeter, Chabrier, and Kroupa.  Finally, \citetalias{BaSTI} is the only model set herein to publish models with alpha enhanced metallicities.  This information is provided as a quick reference for comparing model sets and is summarized in Table \ref{tbl:models} which includes the number of ages in each model set, the number of metallicities provided, and the IMFs provided.

To facilitate direct comparisons between model sets we interpolate between the models to generate a new set of models for each model set with the same metallicities.  Our new models have metallicities of Z = 0.05, 0.1, 0.2, 0.4, 0.8, 1.0, and 1.5 times $Z_\odot$ or Z = 0.001, 0.002, 0.004, 0.008, 0.016, 0.02, and 0.03.  For all of our comparisons below we use these interpolated models.  We also choose to restrict our comparisons to models with the same IMF.  As there is no IMF that is covered by all five model sets we do all comparisons using a Salpeter IMF, and therefore in the comparisons below the models from \citetalias{BaSTI} are not included.  For each of our interpolated models we use EzGal to generate four CSP models.  The CSPs are dust-free, exponentially decaying bursts with e-folding timescales of 0.1, 0.5, 1.0, and 10.0 Gyrs.

\subsection{Filter Set Overview}

For convenience to {\itshape EzGal} users and to enable a basic model comparison we generate a filter set for use with {\itshape EzGal}.  Our filter set includes many commonly used filters: the {\itshape GALEX} FUV and NUV filters, the Sloan filters, all wide {\itshape HST} ACS WFC and WFC3 filters, 2MASS filters, {\itshape Spitzer} IRAC filters, and the {\itshape WISE} 3.4 and 4.6$\mu$m filters.  The filters come from a number of sources and all represent total transmission: CCD, telescope, filter, and a basic atmosphere when appropriate.  The properties of the filter set are summarized in Table \ref{tbl:filters} which has the pivot wavelength and rectangular width for each filter as well as the absolute AB magnitude of the Sun through each filter and the calculated AB to Vega conversion.  The latter is in magnitudes such that the Vega magnitude of a galaxy is its AB magnitude plus the listed conversion.  This Table is also reproduced on the web for quick reference.

\subsection{Comparison}\label{sec:comparison}

\begin{figure}
\epsscale{1.0}
\plotone{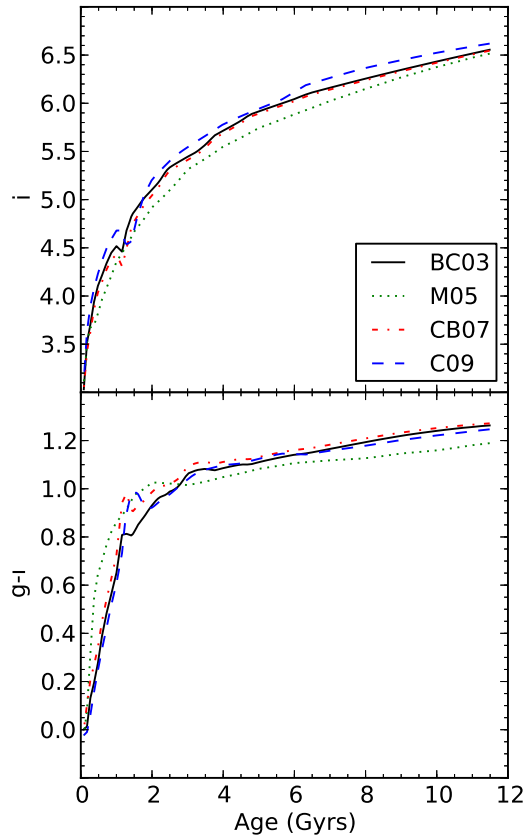}
\caption{A comparison between the predicted i' band rest-frame absolute magnitudes (top) and g'-i' colors (bottom) as a function of age for solar metallicity SSPs with a Salpeter IMF.  The solid, dotted, dash dotted, and dashed lines correspond to the \citetalias{BC03}, \citetalias{M05}, \citetalias{CB07}, and \citetalias{C09} models, respectively.  Magnitudes and colors are on the AB system.}\label{fig:simple_comparison}
\end{figure}

\begin{figure*}
\epsscale{1.0}
\plotone{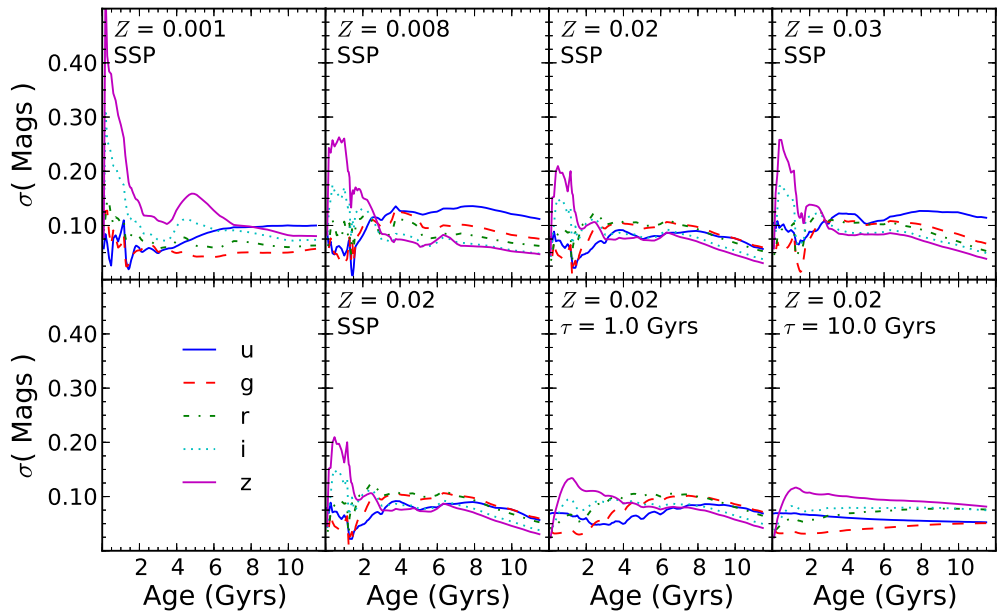}
\caption{Scatter between the predicted magnitudes of the models for the Sloan filters as a function of age.  The solid, dashed, dash-dotted, dotted, and solid lines represent the Sloan u', g', r', i', and z' filters, respectively.  In all panels the standard deviation between the magnitude predictions of four model sets \citepalias{BC03,M05,CB07,C09} through each filter is plotted versus age.  In the top series of panels all the models are for a SSP with a Salpeter IMF and various metallicities: $Z=0.001$ (far left), $Z=0.008$ (left), $Z=0.02$ (right), and $Z=0.03$ (far right).  The bottom series of panels are for stellar populations with a Salpeter IMF and solar metallicity, but for varying star formation histories.  An SSP (left), a dust-free exponentially decaying burst with $\tau = 1.0$ Gyr (right), and a dust-free exponentially decaying burst with $\tau = 10.0$ Gyr (far right).}\label{fig:sloan}
\end{figure*}

We begin our comparison by examining the fidelity of magnitude predictions for the models in Sloan filters.  Figure \ref{fig:simple_comparison} shows the predicted i' band rest-frame absolute magnitude (top) and g'-i' color as a function of age for SSP models with a solar metallicity and Salpeter IMF.  As can be seen from this Figure, differences are typically $\sim$0.1 - 0.2 magnitudes.  To better explore how the scatter depends on age and wavelength, we plot the scatter between models as a function of age, wavelength, metallicity and star formation history in Figure \ref{fig:sloan}.  The top left panel of this Figure illustrates the scatter between the predicted magnitudes of the models (\citetalias{BC03}, \citetalias{M05}, \citetalias{CB07}, and \citetalias{C09}) for the Sloan filters u', g', r', i', z' as a function of age for an SSP with a Salpeter IMF and a metallicity of $Z=0.001$.  The panels to the right show the same thing but for $Z=0.008$, $Z=0.02$, and $Z=0.03$.  The bottom row of panels shows the impact of changing star formation histories.  All the models in the bottom panel have solar metallicity ($Z=0.02$) and a Salpeter IMF.  The first plot on the bottom rows shows the scatter between the model sets for an SSP, the next for a dust-free exponentially decaying burst of star formation with an e-folding time ($\tau$) of 1.0 Gyrs, and the last for a dust-free exponential burst with $\tau = 10.0$ Gyrs.  Scatter in this case refers to the standard deviation of the magnitudes predicted by the different models at a given age and through a particular filter.

A number of conclusions can be drawn from Figure \ref{fig:sloan}.  First, the best-case comparison is for solar metallicities and intermediate to old ages ($\gtrsim$ 4 Gyrs), for which differences between the models are at most 0.1 mags and drop to $\sim$0.05 mags at the oldest ages.  For the Sloan i' and Sloan z' filters the scatter increases by a factor of $\sim$2 for younger ages ($\lesssim$ 2 Gyrs).  This is particularly true for sub-solar metallicities, and the scatter in Sloan i' and Sloan z' increases systematically at these young ages when going from metallicities of $Z=0.02$ to $Z=0.008$ and $Z=0.001$, reaching differences as large as $\sim$0.4 mags.  For the three bluest Sloan filters the scatter is $\lesssim 0.1$ mags for all ages and metallicities.

The bottom series of panels highlights the impact of an extended star formation history, the effect of which is to smooth out the scatter between models as a function of age.  At longer wavelengths when the models differ more at younger ages, this smoothing has a tendency to increase errors at latter times and decrease errors at earlier ones.  Therefore in this case model uncertainty for extended star formation histories will be larger at later times if the star formation history includes a substantial presence of young stars (ages $\lesssim$ 3 Gyrs).

\begin{figure*}
\epsscale{1.0}
\plotone{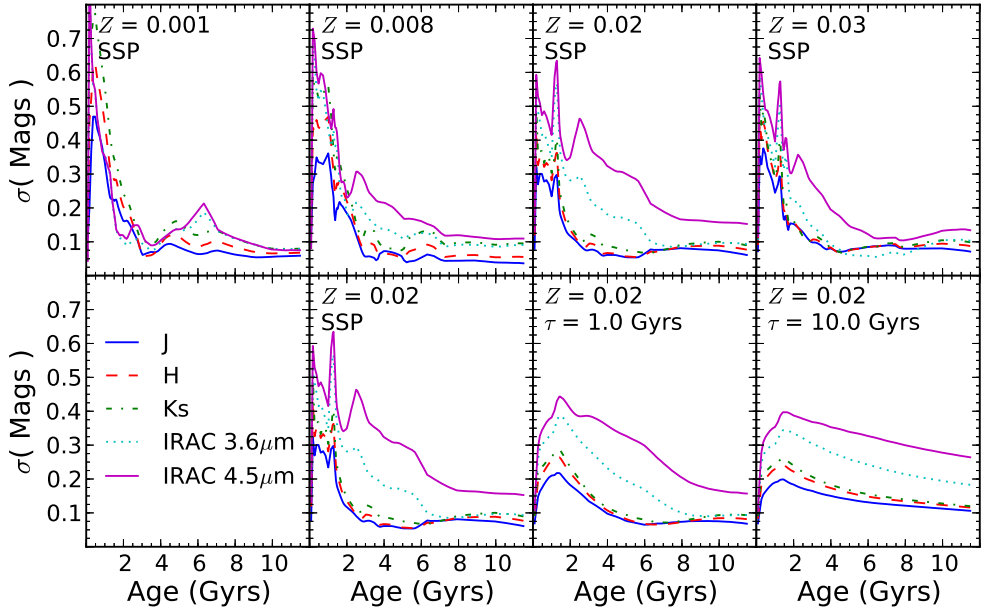}
\caption{Same as Figure \ref{fig:sloan}, but for 2MASS J, H, K$s$ and Spitzer/IRAC 3.6 and 4.5$\mu$m (solid, dashed, dash-dotted, dotted, and solid, repsectively).}\label{fig:ir}
\end{figure*}

Figure \ref{fig:ir} is the same as Figure \ref{fig:sloan} but now various near-IR bands are plotted: J, H, K$s$, and Spitzer/IRAC 3.6 and 4.5$\mu$m.  The first thing to note is that for an older ($\gtrsim$ 3 Gyr) solar metallicity SSP the differences in JHK$s$ are comparable to the differences in the Sloan bands (i.e. Figure \ref{fig:sloan}), while the Spitzer/IRAC bands typically have larger errors in this same regime.  The scatter between the models now has a stronger age dependency, and for ages $\lesssim$2 Gyrs the model uncertainty increases to 0.3 mags (J) and 0.6 mags (3.6$\mu$m).  Metallicity has the opposite impact on the scatter between models in the NIR for young ($\lesssim$ 3 Gyrs) and intermediate to old ($\gtrsim$ 3 Gyrs) stellar populations.  For younger ages the scatter increases systematically while going to lower metallicities.  This effect is particularly pronounced in the K$s$ band which has a maximum scatter of $\sim$0.35 mags for young stellar populations with solar metallicity, but a maximum scatter of $\sim$0.7 mags for young stellar populations with $Z=0.001$.  For older stellar populations the scatter is roughly constant or even decreasing (IRAC 3.6 and 4.5$\mu$m) as the metallicity decreases.

The general trend of increasing scatter towards younger ages is by no means a new discovery but is strongly influenced by uncertainties with the thermally pulsating AGB (TP-AGB) phase (\citetalias{M05}, \citealt{marigo08}, \citetalias{C09}).  This short lived phase in stellar evolution is poorly understood observationally and theoretically; observationally due to its rarity and theoretically because the properties of a TP-AGB star are strongly dependent upon mass loss, which is not predicted theoretically \citep{C09}.  Unfortunately for stellar modeling, TP-AGB stars can dominate the light of a stellar population at long wavelengths ($\lambda \ga 1\mu$m) for ages $\ga 10^8$yrs.  While it is most important in the NIR, it can also impact red optical filters to a smaller extent \citepalias{C09}, and so can readily explain the systematic trend to higher scatters seen as a function of wavelength and age in Figures \ref{fig:sloan} and \ref{fig:ir}.  Moreover, it can exacerbate differences for models with different metallicities because the TP-AGB stars used to calibrate the models typically have unknown metallicities \citep{C09}, creating an additional source of uncertainty.  This likely explains the substantially higher scatter seen for young ages, sub-solar metallicities, and long wavelengths.

The differences seen in Figures \ref{fig:sloan} and \ref{fig:ir} are best viewed as lower limits for the uncertainties introduced by SPS modeling.  This is because agreement between the models can simply be caused by similar methodologies used by the various modeling groups, and does not necessarily imply that the models are doing a better job of agreeing with actual stellar populations.  For instance, we noted above that for old stellar populations the scatter between models is typically the same or smaller for sub-solar metallicities than for solar metallicities.  This fact is not surprising since all the model sets used herein are all compared to or calibrated to match Milky Way globular clusters, which are old and metal poor systems.

\begin{figure*}
\epsscale{1.0}
\plottwo{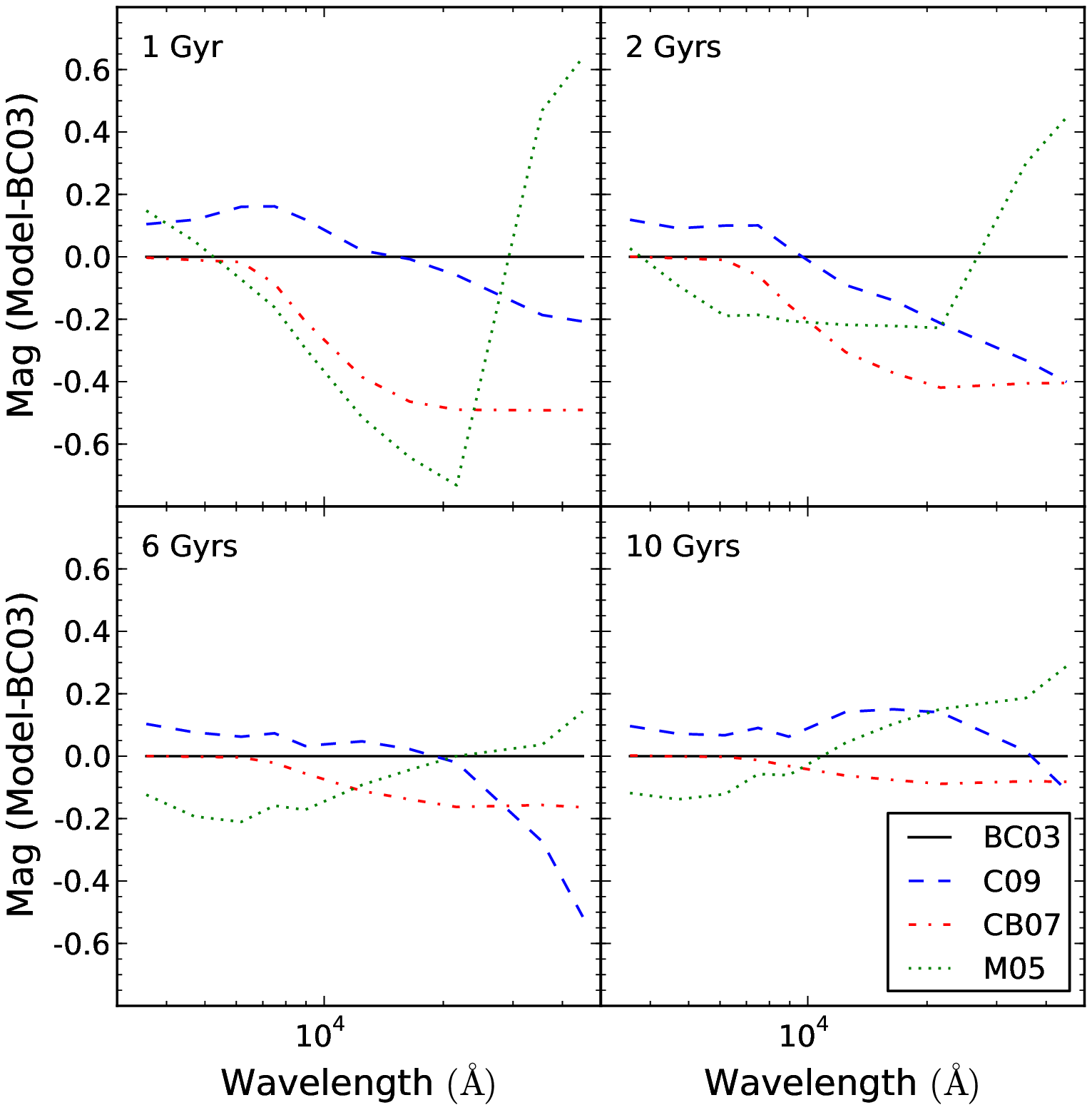}{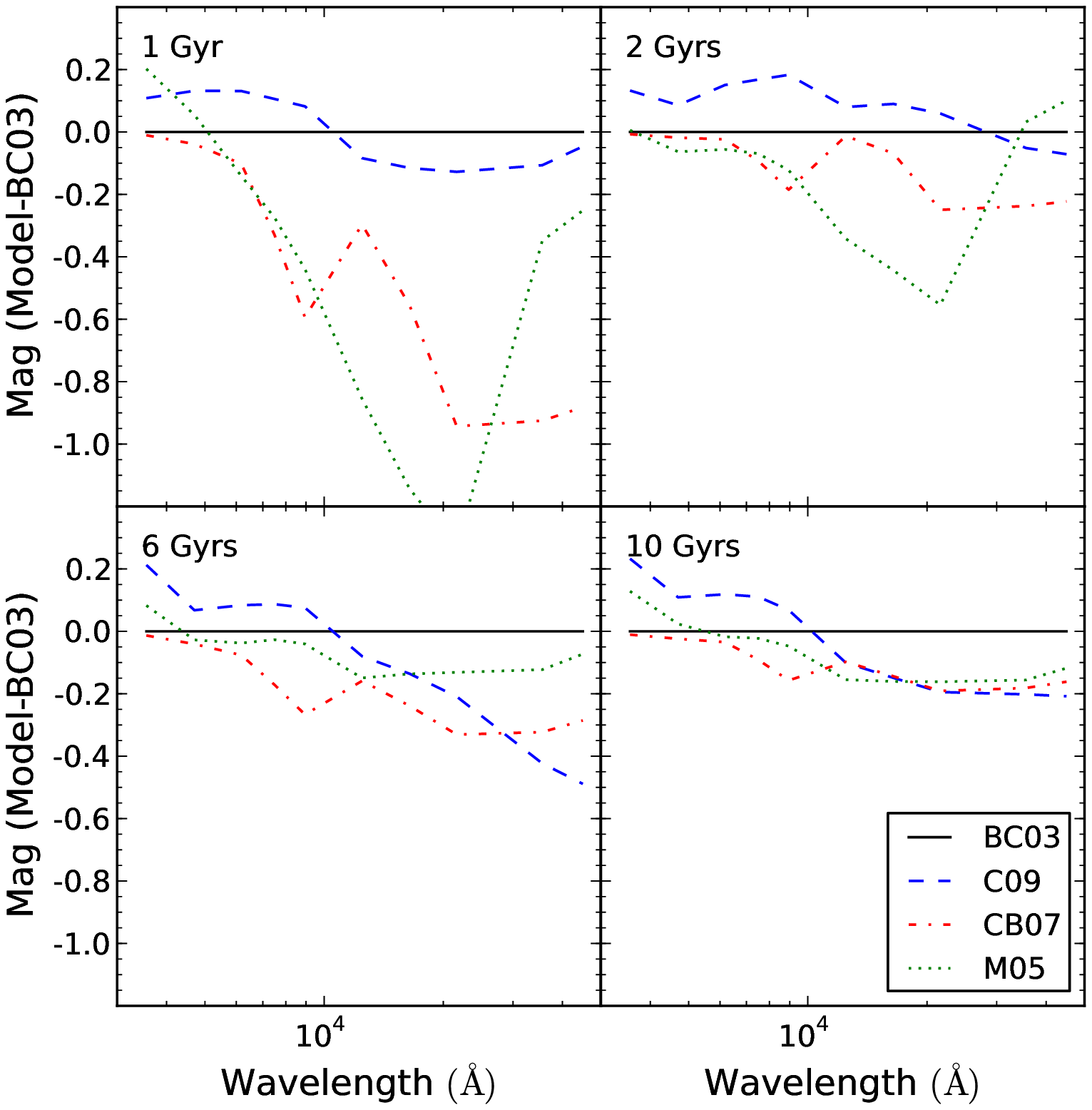}
\caption{Differences between models for solar (left, $Z=0.02$) and subsolar (right, $Z=0.001$) metallicities.  Models are for a SSP with a Salpeter IMF.  Solid, dashed, dash-dotted, and dotted lines represent the difference between the predicted rest-frame absolute magnitudes of \citetalias{BC03}, \citetalias{C09}, \citetalias{CB07}, and \citetalias{M05} (respectively) minus the predictions of the \citetalias{BC03} model set as a function of filter wavelength.  Each panel is divided up into four different plots representing the differences between model predictions at four different ages: 1 Gyr (top left), 2 Gyrs (top right), 6 Gyrs (bottom left), and 10 Gyrs (bottom right).}\label{fig:diffs}
\end{figure*}

Figure \ref{fig:diffs} demonstrates that the scatters seen in Figures \ref{fig:sloan} and \ref{fig:ir} are not driven by just one model set.  This Figure shows the differences between the predicted magnitudes of these four models through the Sloan and NIR filters for four different ages and two metallicities.  All the models in this Figure are SSPs with a Salpeter IMF.  The left panel in Figure \ref{fig:diffs} is for models with solar metallicity and the right panel is for models with a metallicity of $Z=0.001$.  Each panel is divided up into four plots corresponding to four different ages: 1 Gyr (top left), 2 Gyrs (top right), 6 Gyrs (bottom left), and 10 Gyrs (bottom right).  The lines in each plot represent the differences between the predicted absolute magnitude through each filter in each model set \citepalias{BC03,M05,CB07,C09} minus the predicted absolute magnitude of \citetalias{BC03}.  In general, the models are distributed throughout the full range of magnitudes covered by the models.  This shows that apparent disagreements in Figures \ref{fig:sloan} and \ref{fig:ir} are not caused by one discrepant model set.  Therefore, the scatter seen in Figures \ref{fig:sloan} and \ref{fig:ir} is representative of the general uncertainties between the SPS models.

Finally, we note that our results are robust against the choice of model sets used for our comparison.  For instance it might seem expedient to exclude the \citetalias{BC03} models from the above analysis because substantial effort has been put forth to understand the TP-AGB phase since \citetalias{BC03} was published.  However, excluding this model set from the analaysis makes no appreciable differences in our results, which simply reflects the fact that the \citetalias{BC03} models are rarely an outlier in Figure \ref{fig:diffs}.  Our conclusions also remain unchanged if we instead compare the \citetalias{BaSTI}, \citetalias{M05}, and \citetalias{C09} model sets with a Kroupa IMF.  This once again emphasizes that the differences noted in this paper reflect general uncertainties in SPS modeling and are not caused by one discrepant model set.

\subsection{A Practical Example}

We perform one final model comparison to demonstrate the utility of {\itshape EzGal} as well as to reinforce the above results and show how they can impact current work.  We generate new CSP models for our model sets using a more physically motivated SFH, which is the global SFH of the universe.  We use the global star formation rate density as a function of redshift from \citet{gonzalez10} which includes data from \citet{reddy09}, \citet{bouwens08}, \citet{bouwens07}, and \citet{schiminovich05}.  This gives the relative star formation rate in the universe as a function of redshift from $z=0.3$ to $z=8.5$.  We further set the star formation rate to zero at $z=0$ and $z>10$ to prevent our star formation history from having any discontinuities.  While the star formation rate is unlikely to turn on suddenly at $z=10$ or turn off at $z=0$, in practice this assumption makes little difference and does not impact our example.

\begin{figure}
\epsscale{1.0}
\plotone{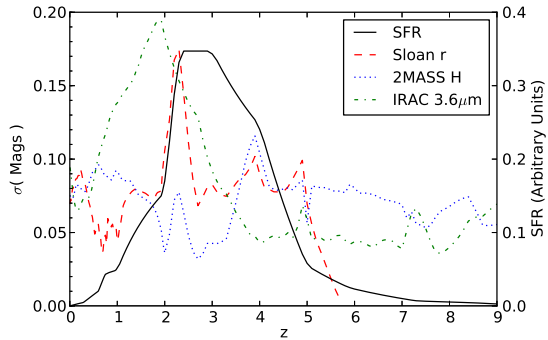}
\caption{Scatter between models as a function of redshift.  The dashed, dotted, and dot-dashed lines show the scatter between predicted apparent magnitudes in the Sloan r', 2MASS H, and IRAC 3.6$\mu$m filters respectively as a function of redshift for \citetalias{BC03}, \citetalias{M05}, \citetalias{CB07}, and \citetalias{C09} models with solar metallicity, a Salpeter IMF, and with a SFH given by the global star formation history of the universe.  The solid line shows the global SFH used (in arbitrary units), which comes from \citet{gonzalez10}.  Scatter refers to the standard deviation of the predicted magnitudes from all 4 bands at a given redshift and for a given filter.  The scatter in Sloan r' cuts off at $z\sim6$ where it is tracing rest-frame wavelengths blueward of the Lyman limit and is therefore unobservable.}\label{fig:z_scatter}
\end{figure}

Using {\itshape EzGal} we generate a CSP from this star formation history for a solar metallicity galaxy with a Salpeter IMF for our four comparison models (\citetalias{BC03}, \citetalias{M05}, \citetalias{CB07}, and \citetalias{C09}).  We then use {\itshape EzGal} to generate apparent magnitude predictions for each CSP model through the Sloan r', 2MASS H, and IRAC 3.6$\mu$m filters as a function of redshift, assuming a formation redshift of $z_f=10.0$.  Finally we calculate the scatter between the predicted magnitudes of the models in the same way as in our previous comparisons.  We show the scatter between models as a function of filter and redshift in Figure \ref{fig:z_scatter}, as well as the star formation history used to generate the CSP models.

The trends seen in Figure \ref{fig:z_scatter} are caused primarily by two effects: increasing model uncertainty for younger stellar populations and the changing rest-frame wavelengths traced by each filter as a function of redshift.  For $3.6\mu$m the model scatter peaks in the $1 \la z \la 3$ range.  At higher redshifts the $3.6\mu$m filter traces the rest-frame optical where the models agree well.  Strong star formation from $2 \la z \la 5$ guarantees that there is a substantial presence of young stars over this redshift range, and therefore the increasing importance of TP-AGB stars leads to increased uncertainty, as does the fact that the $3.6\mu$m filter traces longer wavelengths where TP-AGB stars are again more important.  For $z \la 2$ the star formation rate begins to drop and the stellar populations become steadily older.  Since the models agree well for old ages, this causes the model scatter to peak shortly after the star formation rate peaks and then steadily decline to $z=0$.

In the H-band the scatter between models is relatively constant and typically $\la$0.1 mags.  This low scatter occurs because the H-band filter always traces regions of parameter space for which the models agree well.  At high redshift when the stellar populations are young and the impact of TP-AGB stars is important, the H-band traces the rest-frame optical which is unaffected by TP-AGB stars.  At low redshift the steadily dropping star formation rate leads to an increasing mean age, once again minimizing the impact of TP-AGB stars and leading to low uncertainties.

Similarly for Sloan r' the scatter between models is typically $\la$0.1 mags at low and high redshift.  However, there is a strong and sudden peak in the model scatter at $z\sim2$.  This same feature is also present at precisely the same redshift and significance in all the Sloan filters and the J band, although we only show Sloan r' in Figure \ref{fig:z_scatter}.  The fact that this peak shows up in a variety of filters at the same redshift means that the underlying uncertainty depends primarily on age, not wavelength.  At this redshift the Sloan filters and the J band are all tracing rest-frame wavelengths blueward of the 4000\AA ~break.  In contrast both the H band and IRAC 3.6$\mu$m filters trace wavelengths redward of the 4000\AA ~break, and neither has evidence for a similar increase in model scatter.  Therefore we conclude that this peak in model scatter is caused by uncertainty in modeling young stellar populations blueward of the 4000\AA ~break.

Most importantly, Figure \ref{fig:z_scatter} illustrates one more reason why it is important to use quantitative methods to estimate the impact of SPS model uncertainties.  Observations of galaxies at various redshifts through a given filter will trace stellar populations with a variety of ages and wavelengths.  Moreover, the uncertainties in SPS modeling depend sensitively on wavelength and age.  The result of these facts is that, in practice, SPS model uncertainty often depends on redshift in hard to predict ways.  Therefore, for studies that investigate how stellar populations evolve as a function of redshift it is vital to verify that this redshift dependent model uncertainty is not causing spurious results.  This is best done through quantitative comparison of the models to each other or of the observations to many different models, tasks that {\itshape EzGal} is designed for.

\section{EzGal Web Resources}\label{sec:web}

A number of {\itshape EzGal} resources are available through the {\itshape EzGal} website,\footnote{http://www.baryons.org/ezgal/} including two different interfaces.  The first\footnote{http://www.baryons.org/ezgal/model} allows the user to instantly download magnitude, $k$-correction, $e$-correction, $e+k$-correction, mass-to-light ratio, mass, and solar magnitudes for a given model set and filter as a function of redshift for a set of precalculated formation redshifts and cosmologies.  The second interface\footnote{http://www.baryons.org/ezgal/model\_server} allows for arbitrary choice of formation redshift and cosmology and emails the calculated results to you, which typically takes a minute or two.

An up-to-date copy of Table \ref{tbl:filters} is maintained on the {\itshape EzGal} website with basic filter information, solar magnitudes, and calculated AB to Vega conversions listed for all filters available through the website.  Also distributed with this table is a plot of magnitude, mass-to-light ratio, and $k$-correction evolution as a function of redshift for each filter, a plot of the filter response curve, and a data file giving the filter response curve used by EzGal.

A download page is provided where the source code for {\itshape EzGal} can be downloaded, as well as {\itshape EzGal}-ready model files.  This includes the original SSP models distributed with all the model sets discussed in this paper, as well as the interpolated SSPs and generated CSPs that we use for our comparison.  Finally we distribute a manual for the {\itshape EzGal} API describing how to use {\itshape EzGal} from within python.

\section{Conclusions}\label{sec:conclusions}

{\itshape EzGal} provides a convenient framework for transforming SPS models from theoretical quantities to directly observable magnitudes and colors.  It includes code for generating composite stellar population models with arbitrary star formation histories and dust extinction.  In principle it can work with any model set, providing a simple and consistent framework for comparing the predictions of different model sets and estimating errors introduced by the choice of model set.

We demonstrate the latter property of {\itshape EzGal} by predicting the magnitude evolution for five model sets (\citetalias{BC03}, \citetalias{M05}, \citetalias{CB07}, \citetalias{C09}, \citetalias{BaSTI}) as a function of age, filter, metallicity, and star formation history.  We compare the predictions between the models and note substantial uncertainty (0.3-0.7 mags) for young stellar populations (ages $\la 2$ Gyrs) at long wavelengths ($\lambda \ga$ 1$\mu$m), a region of well-known uncertainty caused by the contribution of thermally pulsating AGB stars.  We note that for old ages, optical filters, and solar metallicities the models agree at the $\sim0.1$ mag level.  For old ages at all wavelengths the models agree as well if not better at sub-solar metallicities than at solar metallicities, which likely reflects the fact that the models are all compared to or calibrated with Milky Way globular clusters.  These differences are best viewed as lower limits on the uncertainties inherent in SPS modeling because it does not include systematic errors in assumptions or methodologies that are shared by all model sets.

Finally we calculate the scatter between our models for a solar metallicity stellar population with a Salpeter IMF and a star formation history matching the global star formation history of the universe.  We conclude that the derived model uncertainty depends upon redshift and filter in hard to predict ways.  This highlights the importance of using quantitative methods to estimate model uncertainty, especially when comparing observations to models as a function of redshift.

These results illustrate the utility of {\itshape EzGal} in simplifying the process of working with SPS model sets, making it easy to compare observations to multiple model sets as well as to compare model sets to each other.  In turn, this provides a simple method to quantify the uncertainties introduced by choice of SPS model set, as well as to find robust or disparate regions in parameter space (age, wavelength, metallicity, etc).

We hope this will help other researchers both in interpreting data and planning new observations.  If you use {\itshape EzGal} in your research we appreciate a citation to this paper.  However, it is especially important that you cite the paper that describes the model set you are using, since {\itshape EzGal} is simply an interface for working with already existing models.  We were not involved in the creation of any of the model sets discussed in this paper, although the authors have been generous enough to allow us to redistribute their models through our web interface.  So when using model sets distributed with {\itshape EzGal} please be sure to cite the appropriate papers.  If you download models through our website, you will find lists of references for the model sets on the download pages.

\acknowledgments

We gratefully acknowledge the authors of all five of the model sets included in this paper for giving us permission to redistribute their work in this way.  We especially thank Charlie Conroy, Maurizio Salaris, Santi Cassisi, St\'{e}fane Charlot, Gustavo Bruzual, and Claudia Maraston for their input on this project.  We are also grateful to our many collaborators - Adam Stanford, Peter Eisenhardt, Yen-Ting Lin, Greg Snyder, and others who have tested {\itshape EzGal} extensively and provided us with invaluable feedback.  Finally, we would like to thank the anonymous referee who's comments have greatly improved the presentation in this paper.  This paper is based upon work supported by the National Science Foundation under grant AST-0708490.

\bibliographystyle{apj}
\bibliography{ms}

\end{document}